\documentclass[runningheads]{llncs}
\usepackage[T1]{fontenc}
\usepackage{graphicx}
\usepackage{latexsym}
\usepackage{graphicx}
\usepackage{amsmath,amsfonts,amssymb}
\usepackage{mathtools}
\usepackage[font=small,labelfont=bf]{caption}
\usepackage{lipsum}
\usepackage{booktabs}
\usepackage{multirow}
\usepackage[table]{xcolor}
\usepackage{arydshln}
\colorlet{myPurple}{blue!60!red}
\usepackage{color, colortbl}
\definecolor{Gray}{gray}{0.9}
\definecolor{ao(english)}{rgb}{0.0, 0.5, 0.0}
\definecolor{cardinal}{rgb}{0.77, 0.12, 0.23}
\usepackage{url}
\usepackage{float}
\usepackage{tabularx}
\usepackage{appendix}
\usepackage{placeins}
\usepackage{enumitem}
\setlist[itemize]{leftmargin=*}
\makeatletter
\newcommand{\ssymbol}[1]{^{\@fnsymbol{#1}}}
\makeatother
\newcolumntype{?}{!{\vrule width 3pt}}
\usepackage{pdflscape}
\usepackage{rotating}
\begin{document}
\title{Multi-Modal Video Topic Segmentation with Dual-Contrastive Domain Adaptation}
\titlerunning{Multi-Modal Video Topic Segmentation}
%
\author{Linzi Xing\thanks{Work done while the first author was an intern at Adobe Research.}\inst{1} \and
Quan Tran\inst{2} \and
Fabian Caba\inst{2} \and
Franck Dernoncourt\inst{2} \and
Seunghyun Yoon\inst{2} \and
Zhaowen Wang\inst{2} \and
Trung Bui\inst{2} \and
Giuseppe Carenini\inst{1}
}
\authorrunning{L. Xing et al.}
%
\institute{University of British Columbia, Vancouver, BC, Canada \and
Adobe Research, San Francisco, CA, USA \\
\email{\{lzxing, carenini\}@cs.ubc.ca}\\
\email{\{qtran, caba, dernonco, syoon, zhawang, bui\}@adobe.com}}
%
\maketitle              
\begin{abstract}
Video topic segmentation unveils the coarse-grained semantic structure underlying videos and is essential for other video understanding tasks. Given the recent surge in multi-modal, relying solely on a single modality is arguably insufficient. On the other hand, prior solutions for similar tasks like video scene/shot segmentation cater to short videos with clear visual shifts but falter for long videos with subtle changes, such as livestreams. In this paper, we
introduce a multi-modal video topic segmenter that utilizes both video transcripts and frames, bolstered by a cross-modal attention mechanism. Furthermore, we propose a dual-contrastive learning framework adhering to the unsupervised domain adaptation paradigm, enhancing our model's adaptability to longer, more semantically complex videos. Experiments on short and long video corpora demonstrate that our proposed solution, significantly surpasses baseline methods in terms of both accuracy and transferability, in both intra- and cross-domain settings.

\end{abstract}
\section{Introduction}
 
\begin{figure*}[t]
    \centering
    \includegraphics[width=0.97\textwidth]{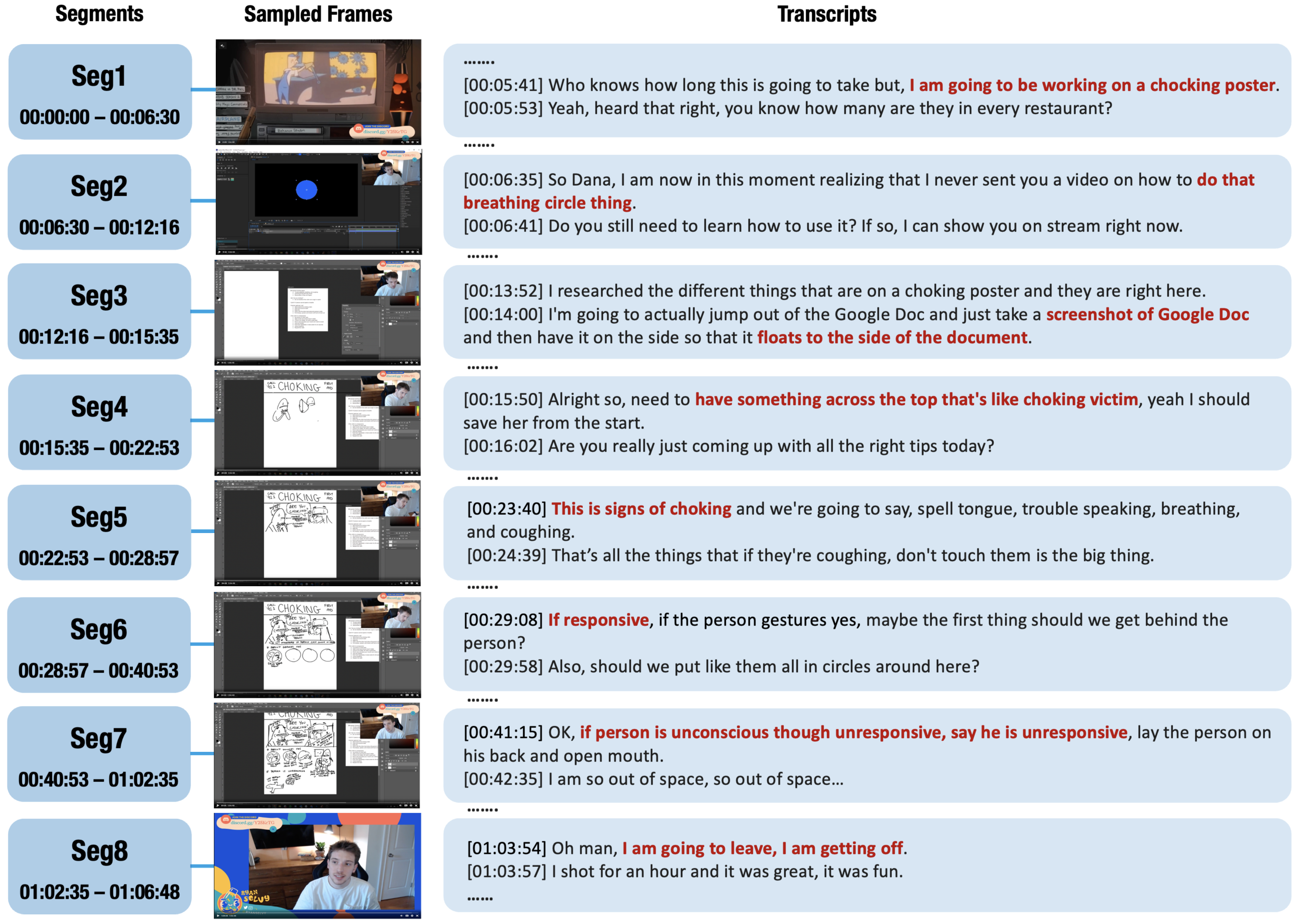}
    \caption{A Behance exemplar about making a choking victim poster. 
    The left side of the figure illustrates the video's timeline after topic segmentation. The right side shows the transcript with words indicating a segment's topic \textbf{\textcolor{cardinal}{highlighed}}.} 
    \label{fig:behance_example}
\end{figure*}

Video Topic Segmentation aims to break stretches of videos into smaller segments consisting of video frames or clips consistently addressing a common topic. As an example given in Figure~\ref{fig:behance_example}, video topic segmentation does the job of segmenting a creative
livestream video (e.g., from livestream platform \textit{Behance\footnote{\url{https://www.behance.net/live}}}) into a sequence of
topical-coherent pieces (\texttt{Seg1} -- \texttt{Seg8}), placing a boundary where a topic transition happens. This task can enhances both human-to-human and human-to-system interactions in modern social contexts, improving real-time engagement on live streaming platform \cite{Faser_chi_20}. In particular, the relatively coarse-grained temporal structure of the input video produced by video topic segmentation is shown to not only simplify video comprehension and helps viewers find content of interest easily. 
More importantly, it can substantially benefit other key video understanding tasks such as video summarization \cite{qiu_etal_2022}, and query-driven video localization \cite{xiao2021boundary}.

Early computational models for video segmentation primarily targeted shot or scene detection by merely leveraging surface visual features like spatiotemporal aspects or frame colors \cite{Rui1998ExploringVS,Rasheed2003SceneDI,rao2020local,Chen2021ShotCS}. These approaches typically measure the temporal similarity along a video's timeline to predict shot/scene boundaries. 
Despite the difference in definition from shot/scene segmentation (discussed in \S\ref{sec:related_work}),
the task of video topic segmentation focuses more on topic-related semantics in the video, which is not necessarily aligned with visual changes. As shown in Figure~\ref{fig:behance_example}, the visual background remains similar in \texttt{Seg3} -- \texttt{Seg7} for a considerable time, even though the streaming topic changes drastically.
Moreover, prior video segmentation methods mostly focused on short videos with clear visual changes and simple patterns \cite{GygliECCV14,slow-dinesh-2016,9250764}.
These distinctive features of short videos could be emphasized in model design or learned in supervised setups, making such models less adaptable to longer, more nuanced video content, like documentaries or instructional livestreams.

To address the aforementioned issues, in this paper, we first propose a simple yet effective multi-modal model for video topic segmentation, which can take both the aligned video transcript and visual frames as input. This ability considerably enhances the model's performance, as textual and visual features can work together to more comprehensively reflect the input video's topic-related semantics \cite{nicolas-etal-combining-2011,Lorenzo-etal-2015}. 
Similar to the formulation used for text topic segmentation \cite{koshorek-etal-2018-text,xing-etal-2020-improving,lo-etal-2021-transformer-pre}, we treat video topic segmentation as a sequence labeling task and introduce a neural model equipped with a cross-modal attention mechanism going beyond simple fusion to effectively integrate textual and visual signals in a complementary manner.
We initially conduct intra-domain experiments by training and testing our proposed segmenter on a newly collected YouTube corpus equipped with high-quality human labeling.
Empirical results show that our multi-modal approach outperforms a set of baseline video segmenters by a substantial margin.

To further adapt our proposed model trained on YouTube videos to longer videos with complex visuals and semantics, we propose an unsupervised domain adaptation strategy empowered by the sliding window inference and dual-contrastive learning scheme. 
Further experiments on two out-of-domain long video corpora demonstrate that the model's generality can be significantly improved 
through applying our dual-contrastive adaptation approach.

\vspace{-1ex}
\section{Related Work} \label{sec:related_work}

\vspace{-1ex}
\paragraph{\textbf{Topic Segmentation}} seeks to uncover the semantic structure of a document (either monologue \cite{xing-etal-2020-improving} or dialogue \cite{xing-carenini-2021-improving}) by dividing it, typically a sequence of sentences, into topical-coherent segments. Recently, a number of supervised neural solutions have been introduced owing to the availability of large-scale labeled corpora sampled from \textit{Wikipedia} \cite{koshorek-etal-2018-text}, with section marks as gold segment boundaries.
Most of these neural segmentation approaches follow the same strategy to simply interpret text segmentation as a sequence labeling problem and further tackle it using a variety of hierarchical neural sequence labelers \cite{koshorek-etal-2018-text,li-etal-2018,xing-etal-2020-improving,xing-etal-2022-improving}.

Inspired by above-mentioned neural text segmenters, our paper similarly frame video topic segmentation as a sequence labeling task due to the availability of textual input (video transcript) and utilize a hierarchical neural sequence labeling framework as the basic architecture of our proposal. Then we extend such framework into the multi-modal setting, with the injection of visual signals from video frames by adding a cross-modal attention network on top of it. More details will be presented in \S\ref{sec:seg_model}.

\vspace{1ex}
\noindent
\textbf{\textit{Shot and Scene Segmentation}} are closely related to video topic segmentation but more narrowly focused. A shot is a sequence of frames from a continuous camera capture. Hence, most shot segmentation techniques mainly use the visual modality to group video frames into shots.
Conversely, a scene is more semantically intricate than a shot, representing a series of related shots that depict events defined by elements like actions, places, and characters. These elements are mostly found in narrative videos such as movies. Thus, past scene segmentation techniques are primarily developed for movies exclusively, aiming to group consecutive shots into scenes based on their visual consistency, spatiotemporal features, or shot color similarity.
Similar but more difficult than shot or scene segmentation, our work focuses on video topic segmentation, which can be deemed as an extension of text topic segmentation with "topic" defined as a relatively self-contained collection of semantically close information.
Notably, the topic's definition relies heavily on the video's context and domain, and isn't strictly bound to the concepts of shots or scenes as previously defined. 
From a machine learning perspective, the more dynamic nature of shot/scene segmentation favours the visual signal, while in our task, the video topic segmentation relies more heavily on the semantic signals carried by natural language \cite{nicolas-etal-combining-2011,Lorenzo-etal-2015}.
Therefore, our video topic segmenter is designed to extend from a topic segmentation framework for text, by integrating visual frames as the auxiliary signal.

\vspace{1ex}
\noindent
\textbf{\textit{Contrastive Learning}} algorithms aim to learn data representations by enlarging the distance between dissimilar samples and meanwhile minimizing the distance between similar samples with contrastive loss functions \cite{Oord2018RepresentationLW}. These techniques have been observed promising in domain adaptation for both uni-modal and cross-modal settings \cite{kang2019contrastive,Kim2021LearningCC,chen2022contrastive}. Recent works on using contrastive learning for video domain adaptation mainly focus on transferring models from source to target domain, where both domains consist of short videos \cite{Kim2021LearningCC}. In contrast, our work attempts to adapt the model trained on short videos to the low-resource domain containing long videos with subtler 
visual changes, by utilizing the semantic overlaps within a single modality or between two modalities to guide the contrastive learning process.

\vspace{-2ex}
\section{Neural Video Topic Segmentation}
\label{sec:seg_model}
\vspace{-1ex}
\subsection{Problem Definition}

Inspired by recent neural-based supervised approaches for text topic segmentation for text, we similarly frame video topic segmentation as a sequence labeling task, with sentences in the video transcript as the units for labeling. More precisely, given an input video with (1) the transcript as a sequence of timestamped sentences, and (2) a sequence of timestamped video frames, our model will predict a binary label for each transcript sentence to indicate whether or not the sentence indicates a topic segment boundary. Formally,

\vspace{1ex}
\noindent
\textbf{\underline{Given}}: A video $v$ with its transcript $T_v$ 
as a sequence of sentences $\{ s_1, s_2, ... , s_n \}$ along 
with start time offsets $\{ b_1, b_2, ... , b_n \}$ / end time offsets $\{ e_1, e_2, ... , e_n \}$, and video frames $X_v = \{ x_1, x_2, ...$ $ , x_m \}$, as a single frame $x_i$ has timestamp $t_i$.

\vspace{1ex}
\noindent
\textbf{\underline{Predict}}: A sequence of labels $\{ l_1, l_2, ... , l_{n-1} \}$ for the sequence of transcript sentences, where $l$ is a binary label, $1$ means the corresponding sentence overlaps a video topic segment boundary, 
$0$ otherwise. We do not predict the label for the last sentence $s_n$, as it is by definition equal to 1, i.e., the end of the last segment.

\begin{figure*}
\setlength{\belowcaptionskip}{-10pt}
    \centering
    \includegraphics[width=0.92\linewidth]{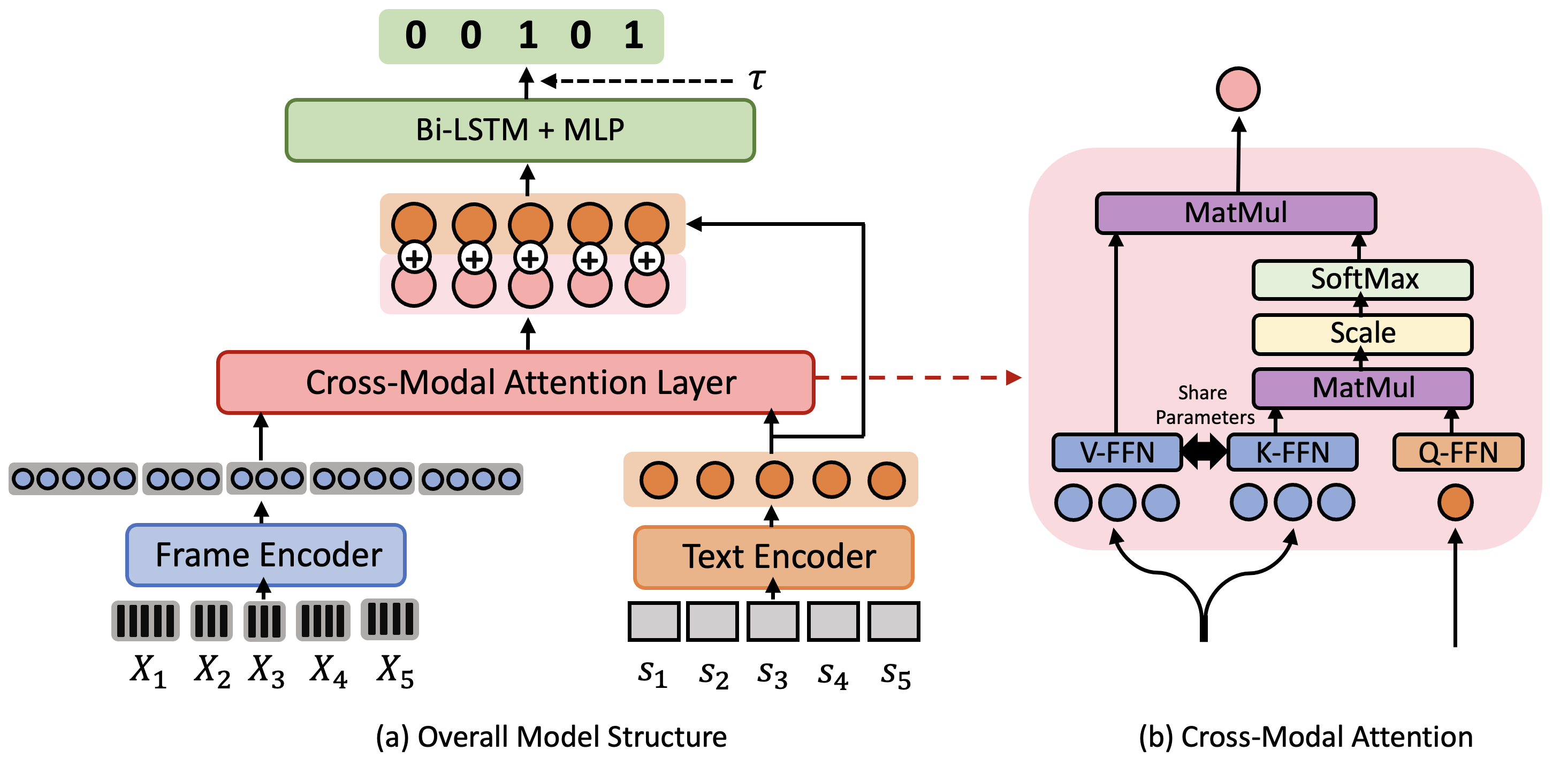}
    \caption{\textbf{Left}: The overall architecture of our proposed multi-modal video topic segmentation model, with four sub-modules (in \S\ref{sec:model_details}) coded in different colors. \textbf{Right}: The detailed illustration of cross-modal attention.} 
    \label{fig:proposed_model}
\end{figure*}

\vspace{-2ex}
\subsection{Model Architecture}
\label{sec:model_details}
Figure~\ref{fig:proposed_model} illustrates the detailed framework of our proposed video topic segmenter, which is similar in architecture to TextSeg \cite{koshorek-etal-2018-text}. It comprises 
two hierarchically linked encoding layers: one as text encoder for contextualized encoding within a sentence (orange in Figure~\ref{fig:proposed_model}) and the other for contextualized encoding between sequence units (green in Figure~\ref{fig:proposed_model}). To allow both textual and visual modalities contribute complementarily to the model's prediction, we add a frame encoder module (blue in Figure~\ref{fig:proposed_model}) and a cross-modal attention mechanism (red in Figure~\ref{fig:proposed_model}). This design choice extends the framework to the multi-modal setting, accepting now both the video transcript and frames as input and making prediction of topic segment boundaries based on them.

The \textbf{\texttt{text encoder}} module $E_t$ yields low-level features for sentences in the video transcript. Different from the proposal in \cite{koshorek-etal-2018-text} using the BiLSTM + attention as the backbone for text encoder, here we adopt the pre-trained vanilla BERT \cite{devlin-etal-2019-bert} in accordance to its achieved superiority on text segmentation observed in \cite{xing-etal-2020-improving} and \cite{lukasik-etal-2020-text}. Parallel to the text encoder, a \textbf{\texttt{frame encoder}} $E_f$ is introduced to extract features for visual signals with the standard pretrained ResNet-18 
\cite{7780459}. Formally, given a transcript sentence $s_i$ with its time interval as $[b_i, e_i]$ and a set of video frames $X_i = \{ x^i_1, ... , x^i_m \}$ associated with this sentence, where each frame (e.g., the $k^{th}$ frame) in the set has timestamp $t^i_k \in [b_i, e_i]$, we can obtain the sentence representation $tr_i = E_t(s_i)$ and its corresponding set of frame representations \textit{FR}$_i = \{ fr^i_1, ... , fr^i_m \}$, where $fr^i_k = E_f(x^i_k)$.

Next, we propose to use a \textbf{\texttt{cross-modal attention}} mechanism to produce a text-aware visual representation for each sentence, rather than obtaining the visual representation by naively operating mean-pooling 
over the set of frame representations covered by the sentence interval.
This design is motivated by the observation that transcript sentences and video frames are in an one-to-many relation, and frames sharing more semantics with the text should be given more attention weights for visual representation generation. Irrelevant frames sharing no or little semantics with the text may negatively affect the quality of the fused multi-modal representation passed to the subsequent module \cite{wang-etal-2022-contrastive-video}.
In practice, the cross-modal attention module adopts the standard scaled dot-product attention function proposed in \cite{NIPS2017_3f5ee243}. With the transcript sentence representation $tr_i$ and its corresponding frame representation set \textit{FR}$_i = \{ fr^i_1, ... , fr^i_m \}$ as query and key (value), we compute the text-aware visual representation ${vr}_i$ as:
\vspace{-1ex}
\begin{equation}
    {vr}_i =  A_i V_i, \label{eq:1}
\end{equation}
\vspace{-2ex}
\begin{equation}
    A_i = softmax(\frac{q_{i}K_{i}^T}{\sqrt{d_k}}) \label{eq:2}
\end{equation}
where $q_i \in \mathbb{R}^{1 \times d_k}$, $K_i \in \mathbb{R}^{m \times d_k}$, and $V_i \in \mathbb{R}^{m \times d_k}$ denote the query vector, key and value matrices generated by passing the sentence representation and frame representations through three parallel feedfoward layers, namely \textit{Q-FFN}, \textit{K-FFN} and \textit{V-FFN} respectively. More formally, we have $q_i = $ \textit{Q-FFN}$(tr_i)$, $K_i = $ \textit{K-FFN}$($\textit{FR}$_i)$, $V_i = $ \textit{V-FFN}$($\textit{FR}$_i)$, where \textit{K-FFN} and \textit{V-FNN} share the same parameters and thus produce identical key and value matrices following \cite{10.5555/3157096.3157129,DBLP:journals/corr/abs-2004-12238}.

Then all the obtained text-aware visual representations $\{ vr_1, ..., vr_{n} \}$ are concatenated with their corresponding sentence representations $\{ tr_1, ..., tr_{n} \}$ and fed into a BiLSTM layer which performs contextualization and returns hidden states. Next, a multilayer perceptron (MLP) followed by Softmax serves as a topic boundary predictor to make binary predictions regarding the input hidden states according to a threshold $\tau$ tuned on the validation set. More specifically, if a transcript sentence's output probability exceeds $\tau$, it's marked as $1$, indicating a segment boundary. The entire model is fine-tuned using cross-entropy loss.

We train and test this model on a newly collected YouTube corpus (in \S\ref{sec:datasets}) and empirically verify its in-domain effectiveness (reported in Table~\ref{tab:res_intra}).
We leverage it as the source model in the next section to help deliver the domain adaptation strategy coupled with our proposed multi-modal framework. 

\vspace{-1ex}
\section{Long Video Adaptation}
\label{sec:dual_cl}

The rise of (live-)streaming platforms has increased the demand for topical segmentation of videos on these platforms. Unlike videos (e.g., on YouTube) with careful pre-editing and segment labeling provided by their creators, videos on (live-)streaming platforms (e.g., \textit{BBC} documentaries and creative livestreams on \textit{Behance}) are usually extremely long, with sparse visual changes, and more importantly, time-consuming to obtain segment annotations. Thus, it is impractical to learn a fully-supervised model on such videos, and the segmentation model described in \S\ref{sec:model_details} (source model) trained on short YouTube videos (source domain) might underperform when applied to these extensively long videos (target domain).
To adapt the source model for the above-described target domain, 
we propose to equip it with two strategies, namely \textbf{\texttt{Sliding Window Inference}} (\S\ref{sec:slide_window}) and \textbf{\texttt{Dual-Contrastive Adaptation}} (\S\ref{sec:dual_contrast}).

\vspace{-2ex}
\subsection{Simple Sliding Window Inference}
\label{sec:slide_window}

Due to the length discrepancy between videos from the source and target domains, 
directly applying the source model to the target input taking the full transcript sentence sequence as input 
is observed to cause extremely sparse boundary predictions \cite{glavas-2020-two}. Therefore, we propose to calibrate the input length by first breaking the long target input into snippets by a sliding window with the size consistent with the source input length (i.e., the average length of the YouTube videos). 
After applying the source model on every snippet and aggregating outputs of all snippets, we eventually make binary prediction for each sentence if its aggregated probability exceeds the threshold $\tau$ pre-tuned on the source domain. Formally, given a long video input from the target domain with length $= n$ and a fixed window size $= k$ ($n \gg k$),  we can create $n-k+1$ snippets $\{\mathbb{S}_1, ..., \mathbb{S}_{n-k+1} \}$ where each snippet consists of $k$ consecutive sentences by sliding the window over the input with the stride of 1. As a result, each transcript sentence $s_m$ can be covered by up to $k$ snippets. Once we apply the source model to all snippets, we can obtain multiple probability predictions for $s_m$. We then aggregate these probability predictions associated with $s_m$ by taking average:

\vspace{-3 mm}
\begin{equation}
    \bar{p}_m =  \frac{1}{k}\sum_{i=1}^k p_m^{\mathbb{S}_{i}} \label{eq:3}
\end{equation}
Finally, we predict that $s_m$ falls on a segment boundary if $\bar{p}_m > \tau$.

\vspace{-2ex}
\subsection{Dual-Contrastive Adaptation}
\label{sec:dual_contrast}
To transfer the pre-trained source model to the target domain while still preserving its performance on the source domain, 
we follow a more sophisticated unsupervised domain adaptation paradigm, which leverages both labeled source data and unlabeled target data. 
Specifically, we fix the frame and text encoder ($E_t$ and $E_f$) 
while updating the rest of the model in two steps, where the first step updates the model on unlabeled target data using two contrastive learning objectives namely \textit{intra-modal contrastive loss} and \textit{cross-modal contrastive loss}, while the second step updates the model on source data with the supervised training scheme described in \S\ref{sec:model_details}.
 
As the overview of the first step shown in Figure~\ref{fig:domain_adaptation}, given a collection of the paired sentence representations and their corresponding frame representation sets $\{tr_i, $\textit{FR}$_i\}_{i=1}^{b}$ in a training batch (size=$b$) produced by the text and frame encoder from the target domain, 
we need two distinct projection heads to map these sentence/frame representations 
into a shared space. Here we use \textit{Q-FFN} and \textit{K-FNN} (which shares parameters with \textit{V-FNN}) in the cross-modal attention to serve as projection heads for textual and visual modality respectively. Thus we have:
\vspace{-2 mm}
\begin{equation}
    q_i = \textit{Q-FFN(}tr_i\textit{)}, K_i = \textit{K-FFN(}FR_i\textit{)} \label{eq:4}
\end{equation}
where $q_i$ and $K_i=\{ k^i_1, ... , k^i_m\}$ denote the projected sentence embedding and 
the set of projected frame embeddings covered by the sentence. As frames covered by the same sentence are more likely to share similar semantics, we pull the frames attached to the same sentence closer and push the ones from different sentences far apart, by minimizing the intra-modal contrastive loss for visual modality defined as:
\vspace{-4mm}
\begin{equation}
    l^{intra} = -\sum_{i=1}^b log\frac{exp(\tilde{k}^i \cdot \tilde{k}^i)/\tau}{\sum_{j=1}^b exp(\tilde{k}^i \cdot \tilde{k}^j)/\tau} \label{eq:5}
\end{equation}
where $\tilde{k}^i$$\in_R$ \textit{FR}$_i$ and $\tau$ is the hyper-parameter of temperature. To learn the semantic relation between modalities in the target domain, we first obtain the visual representation matched with a sentence by averaging all frames covered by the sentence, and then bring semantically close sentence-visual pairs together and push away non-related pairs by minimizing the cross-modal contrastive loss:
\vspace{-3 mm}
\begin{equation}
    l^{cross} = -\sum_{i=1}^b log\frac{exp(q_i \cdot \textit{MP}(K_i))/\tau}{\sum_{j=1}^b exp(q_i \cdot \textit{MP}(K_j))/\tau} \label{eq:6}
\end{equation}
where \textit{MP} denotes \textit{Mean-Pooling}. The total dual-contrastive loss is formed as:
\begin{equation}
    l = l^{intra} + l^{cross} \label{eq:7}
\end{equation}

\noindent
The above dual-contrastive learning phase is followed by the second step. 
In this step, the model is further trained on the labeled source data again to leverage the target-domain signals learned by the parameters in cross-modal attention for boundary prediction. This step ensures the model adapted to the target domain still preserves some level of effectiveness when applied to the source domain. 

\vspace{2ex}
\noindent
\begin{minipage}{.45\textwidth}
\setlength{\belowcaptionskip}{-5pt}
    \centering
    \includegraphics[width=2.3in]{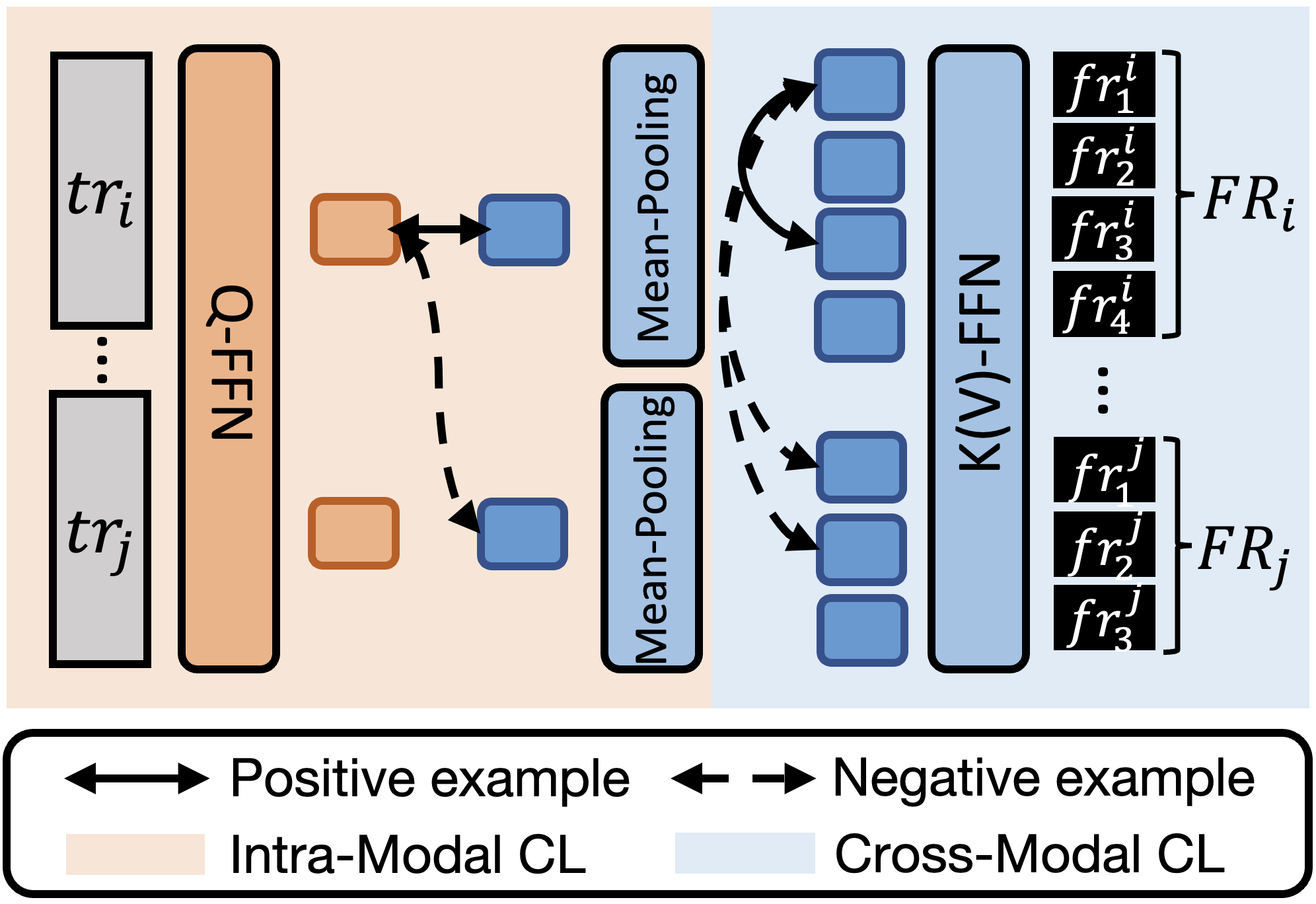}
    \captionof{figure}{An illustration of updating cross-modal attention module with dual-contrastive adaptation.}
    \label{fig:domain_adaptation}
\end{minipage}%
\hfill
\begin{minipage}{.45\textwidth}
    \setlength{\belowcaptionskip}{5pt}
    \centering
    \scalebox{0.76}{
    \begin{tabular}{| l | c c c |}
        \hline
        \rowcolor{Gray}
        \textbf{Dataset} & \textbf{YouTube} & \textbf{BBC} & \textbf{Behance} \\
        \hline
        \# of Vids. & 5,422 & 11 & 575 \\
        \# of Tokens/Vid. & 1,411 & 2,868 & 11,554\\
        \# of Sents./Vid. & 108 & 216 & 1,287 \\
        \# of Segs./Vid. & 6.7 & 29.1 & 5.2 \\
        \# of Tokens/Seg. & 209 & 99 & 2,229 \\
        \# of Sents./Seg. & 16 & 8 & 248 \\
        \# of Frames/Sec. & 10 & 4 & 4 \\
        Avg. Length & 0:09:30 & 0:45:02 & 2:07:51 \\
        \hline
    \end{tabular}}
    \captionof{table}{Statistics of corpora used for training and evaluation in \S\ref{sec:exp}.}
    \label{tab:datasets}
\end{minipage}

\section{Experimental Setup}
\label{sec:exp}
To evaluate the effectiveness and generality of our proposed video topic segmentation model (in \S\ref{sec:seg_model}) and long video adaptation strategy (in \S\ref{sec:dual_cl}), we conduct experiments on two different 
settings, namely \textbf{\textit{Intra-Domain Segment Inference}} and \textbf{\textit{Cross-Domain Segment Inference}}. 

\vspace{-2ex}
\subsection{Intra-Domain Dataset -- YouTube} \label{sec:datasets}
For intra-domain segment inference,  we train and test models with the data from the same domain (corpus). Due to the lack of large-scale human-annotated dataset in the field of video segmentation, 
we collect a novel corpus consisting of 5,422 user-generated videos randomly sampled from YouTube. 
During the video collecting process, we applied filtering criteria to eliminate unsuitable samples to ensure the quality of the dataset to construct. These criteria include constraints on video length ($>$ 100s), word count ($>$ 0.5 word/second on average), chapter durations (all chapters with length $>$ 10s), and sentence length (all sentences with length $<$ 60 tokens) inspired by \cite{cao2022multimodal}.
Each video in this corpus is associated with a series of user-defined chapter timestamps indicating the beginning of each topic chapter contained in this video. We thereby use these available chapter beginning timestamps as ground-truth topic segment boundaries since a topic chapter is by definition a main thematic division within a video. The video transcripts are generated by YouTube ASR with token-level offsets and we further exploit a top performing punctuation restoration model \cite{alam-etal-2020-punctuation} to boost the quality of transcript sentence segmentation. Compared with other existing video understanding corpora such as \textit{BBC} \cite{Lorenzo-etal-2015} (for scene segmentation, size = 11) and \textit{TVSum} \cite{7299154} (for video summarization, size = 50), our constructed YouTube corpus is (1) larger in size; (2) covering more diverse topics; (3) with reliable segmentation which has already been specified by video creators. We split this corpus into train/dev/test portions with size: 5148/134/140.

\vspace{-2.5ex}
\subsection{Datasets for Cross-Domain Inference}

To evaluate our proposal's robustness in cases where a domain-shift is present, 
we conduct experiments for cross-domain segment inference, in which our proposed supervised segmenter is initially trained on the 
YouTube video corpus, and tested on two corpora 
with videos significantly longer than YouTube videos:

\vspace{0.5ex}
\noindent
\textbf{BBC Planet Earth} \cite{Lorenzo-etal-2015} consists of 11 episodes from the BBC educational TV series Planet Earth. Topic segment boundaries of the dataset have been manually annotated by human experts and transcript sentences are obtained using \textit{Whisper}\footnote{\url{https://openai.com/research/whisper}} \cite{radford2022robust}. 
As the statistics in Table~\ref{tab:datasets} indicates, this dataset has longer videos with more segments than YouTube videos, while each segment covers much fewer sentences and tokens.
We split this dataset into 5 and 6 for validation and testing.

\vspace{0.5ex}
\noindent
\textbf{Behance Livestream} consists of 575 videos sampled from the creative livestream 
platform \textit{Behance}. Livestreams, in general, are very challenging to segment into 
coherent sections, since they contain mixes of multi-user dialogues, with 
visual features that change little over the video. They are also very long, with 
hours of content in each video. Behance livestreams, in particular, contain another layer of challenge, 
since they are highly specialized in creative tasks such as animation and image 
edit with unknown entities and intricate visual operations. We believe finding structures in such noisy environment would be not only a good 
benchmark for future topic segmentation research, but also of great practical 
value for consumers, who would find it hard to navigate contents in a back-and-forth 
conversation. Thus, we collect such a corpus and annotate its videos with topic segments given the help from domain experts in animation 
and image edits. For each video, we pay \$23.16 for human annotators to watch and create segments based on the video content. The total cost to annotate 575 videos is \$13,317. We split this dataset into validation and test sets with 57 (10\%), 518 (90\%).

\subsection{Baselines}
\vspace{-1ex}
We consider 
the following representative baselines as comparisons:\\
\vspace{-2ex} \\
\textbf{- Random}: Given a video with its transcript consisting of $k$ sentences, we first randomly sample the segment number $b \in \{0,...,k-1\}$ for this video, and then determine if a sentence overlaps a segment boundary with probability $\frac{b}{k}$. \\
\vspace{-2ex} \\
\textbf{- BayesSeg} \cite{eisenstein-barzilay-2008-bayesian} originally proposed for text segmentation by predicting segment boundaries through modeling the lexical cohesion in a Bayesian context. \\
\vspace{-2ex} \\
\textbf{- GraphSeg} \cite{glavas-etal-2016-unsupervised} generates a semantic relatedness graph with sentences as nodes. Segments are then predicted as the maximal cliques in graph.\\
\vspace{-2ex} \\
\textbf{- Cross-BERT} \cite{lukasik-etal-2020-text} is a supervised text topic segmenter representing candidate segment boundaries using their left and right contexts encoded with BERT. 
The model is trained on YouTube and applied to longer BBC and Behance videos.\\
\vspace{-2ex} \\
\textbf{- TransNet} \cite{soucek2020transnetv2}: This model adopts stacked Dilated DCNN blocks as its basic framework and targets 
video segmentation on shot level. Here we use its publicly available version trained on synthesized video corpora off-the-shelf.\\
\vspace{-2ex} \\
\textbf{- LGSS} \cite{rao2020local} 
is a multi-modal movie scene segmenter pre-trained on the limited-scale \textit{MovieScenes} corpus. 
We use its 
public version with visual frames as input.\\
\vspace{-2ex} \\
\textbf{- X-Tiling} 
is our extension of \textit{TextTiling} \cite{hearst-1997-text}. 
While TextTiling 
only allows textual embeddings as input, X-Tiling 
accepts both textual and visual embeddings and their concatenations as input to compute semantic coherence  and then make segment boundary predictions for videos. 
For fair comparison, the input textual and visual embeddings are produced by pre-trained BERT and mean-pooling over ResNet-18 respectively.

All hyper-parameters required by the above baselines are tuned on the validation portion of the datasets included in this paper.

\vspace{-1ex}
\subsection{Evaluation Metrics}
We apply three standard metrics in previous literature to evaluate the performances of our proposal and baselines. They are:\\
\vspace{-2ex} \\
\textbf{-} $\boldsymbol{F_1}$, with higher scores denoting better performance. It measures the exact match between ground truth and model's prediction.\\
\vspace{-2ex} \\
\textbf{- $\boldsymbol{P_r}$ error score} \cite{georgescul-etal-2006-analysis}, which fixes the inadequacies 
of $P_k$ \cite{Beeferman1999} and \textit{WindowDiff} \cite{pevzner-hearst-2002-critique}, previously considered as two standard evaluation metrics for text segmentation. Concretely, $P_r$ is the mean of missing and false alarm probabilities, calculated based on the overlap between ground-truth segments and model's predictions within a certain size sliding window. Since it is a penalty metric, lower score indicates better performance. \\
\vspace{-2ex} \\
\textbf{- \textit{mIoU}} score \cite{zhu-etal-2022-end}, shortened from the mean Intersection-over-Union. It is calculated by taking average over maximal IoUs of all ground-truth segments to predicted segments. Higher score indicates better performance.

\vspace{-2ex}
\subsection{Implementation Details}
In our multi-modal video topic segmenter, we employ the \texttt{[CLS]} token representation from \texttt{bert-base-uncased} (dimension $d = 768$) for sentence representation and ResNet-18's avg-pooling layer output (dimension $d = 512$) for frame representation. The cross-modal attention's feedforward layers have an output dimension of 768. Our BiLSTM has 2 layers with a hidden size of 256. For YouTube training, we use the Adam optimizer with a learning rate of $1e^{-3}$ and a batch size of 16. In the long video adaptation with dual-contrastive learning, the SGD optimizer is applied with a mini-batch size of 256, learning rate of $3e^{-2}$, and softmax temperature of $1e^{-1}$. Training spans 10 epochs for both supervised learning and domain adaptation, with results averaged over 3 runs. The segmentation threshold, $\tau$, is tuned on validation sets for both intra-domain and cross-domain evaluations.

\vspace{-2ex}

\begin{table}[t]
\setlength{\belowcaptionskip}{-10pt}
\setlength{\tabcolsep}{12pt} 
\centering
\scalebox{0.95}{
\begin{tabular}{l|c|ccc}

\specialrule{.1em}{.05em}{.05em}
\rowcolor{Gray}
\multicolumn{1}{l}{\textbf{Method}} & \multicolumn{1}{c}{\textbf{Modalities}} & \textbf{Pr} $\downarrow$ & \textbf{F1} $\uparrow$ & \textbf{mIoU} $\uparrow$ \\
\cline{1-5}
Random & -- & 45.84 & 48.98 & 37.76 \\
X-Tiling & Text & 39.94 & 51.56 & 50.97 \\
BayesSeg & Text & 40.70 & 50.24 & 49.69 \\
GraphSeg & Text & 38.12 & 51.41 & 51.73 \\
Cross-BERT$^*$ & Text & \underline{32.89} & \underline{60.48} & \underline{60.00} \\
X-Tiling & Visual & 37.53 & 52.08 & 53.24 \\
TransNet & Visual & 40.14 & 51.01 & 50.66 \\
LGSS & Visual & 39.77 & 50.87 & 51.13 \\
X-Tiling & Text + Visual & 38.78 & 52.30 & 51.45 \\
\cline{1-5}
\multirow{3}{*}{\shortstack{NeuralSeg$^*$ \\ (Ours)}} & Text & 31.91 & 63.23 & 61.36 \\
                              & Visual & 50.18 & 47.59 & 16.15 \\
                              & Text + Visual & \textbf{30.61} & \textbf{65.29} & \textbf{63.11} \\

\specialrule{.1em}{.05em}{.05em}
\end{tabular}
}
\caption{\label{tab:res_intra} Results on YouTube for intra-domain evaluation. \textbf{Bold} results indicate the best performance across all comparisons. Underlined results indicate the best performance within their own sub-section. * indicates a fully supervised setting.}
\end{table}

\section{Results and Discussion}
\label{sec:results}

\textbf{Intra-Domain Segment Inference: }Table~\ref{tab:res_intra} reports the performance of the chosen
baselines and our proposal (NeuralSeg in the table) on the YouTube testing set, while
NeuralSeg is trained on the YouTube training set with different input modality settings.
Notably, our model significantly outperforms the best baseline, Cross-BERT, even when only using video transcripts (text) as input. But on the other hand, if we train the model by using only video frames (visual) as input, the model’s performance is even considerably worse than the random baseline, possibly because that too diverse visual and topic presence in the corpus makes it difficult for the model to learn a meaningful visual input-to-prediction
projection. Yet, by combining both modalities with cross-modal attention fusion, the model’s performance can be further enhanced compared with only textual
modality. These results confirm that the visual information itself may not be sufficient to capture a video’s topic-related semantics under a supervised setting, but fusing it
together with the textual information can provide a more clear picture of the video’s
underlying topics.

\begin{table*}
\setlength{\belowcaptionskip}{-10pt}
\setlength{\tabcolsep}{4pt} 
\centering
\scalebox{0.77}{
\begin{tabular}{l |  c  |  c @{\space\space\space\space} c  c | c @{\space\space\space\space} c  c }
\specialrule{.1em}{.05em}{.05em}
\rowcolor{Gray}
\multicolumn{1}{c}{\textbf{Method}} & \multicolumn{1}{c}{\textbf{Modalities}} & \multicolumn{3}{c}{\textbf{BBC}} & \multicolumn{3}{c}{\textbf{Behance}} \\
\hline
  &  & \hspace{1ex} \textbf{Pr} $\downarrow$ & \hspace{1.5ex} \textbf{F1} $\uparrow$ & \hspace{0.8ex} \textbf{mIoU} $\uparrow$ & \hspace{1ex} \textbf{Pr} $\downarrow$ & \hspace{1.5ex} \textbf{F1} $\uparrow$ & \hspace{0.7ex} \textbf{mIoU} $\uparrow$ \\
\hline
Random & -- & 48.29 & 47.36 & 30.66 & 46.27 & 50.00 & 36.24 \\
\hline
X-Tiling \cite{hearst-1997-text} & Text &  \underline{41.59} & 50.72 & 33.27 & 45.93 & 49.95 & \underline{43.16} \\
BayesSeg \cite{eisenstein-barzilay-2008-bayesian} & Text & 42.01 & \underline{51.01} & \underline{33.79} & 49.01 & 48.16 & 37.55 \\
GraphSeg \cite{glavas-etal-2016-unsupervised} & Text & 45.11 & 49.28 & 30.23 & 46.76 & 50.12 & 40.38 \\
Cross-BERT \cite{lukasik-etal-2020-text} & Text & 44.63 & 49.70 & 31.51 & 46.15 & 49.88 & 41.02 \\
X-Tiling \cite{hearst-1997-text} & Visual & 44.48 & 49.89 & 33.26 & \underline{44.56} & \underline{50.73} & 41.81 \\
TransNet \cite{soucek2020transnetv2} & Visual & 42.54 & 49.82 & 31.12 & 46.34 & 49.88 & 40.54 \\
LGSS \cite{rao2020local} & Visual & 42.88 & 50.02 & 32.66 & 45.67 & 50.12 & 39.90 \\
X-Tiling \cite{hearst-1997-text} & Text + Visual & 43.22 & 50.44 & 32.68 & 45.87 & 50.01 & 41.72 \\
\hline
Ours (Plain) & Text + Visual & 43.14 & 54.13 & 31.23 & 45.71 & 51.25 & 34.33 \\
Ours (Window) & Text + Visual & 40.66 & 54.92 & 37.50 & 43.03 & 51.83 & 47.66 \\
\cdashline{1-8}
Ours (Window + CL-Cross) & Text + Visual & \hspace{0.8ex}36.45\textsuperscript{\textdagger} & 54.91 & \hspace{0.8ex}50.92\textsuperscript{\textdagger} & 42.72 & 51.68 & \hspace{0.8ex}48.65\textsuperscript{\textdagger} \\
Ours (Window + CL-Intra) & Text + Visual & \hspace{0.8ex}37.14\textsuperscript{\textdagger} & 55.10 & \hspace{0.8ex}49.22\textsuperscript{\textdagger} & 42.56 & 51.84 & 48.39 \\
Ours (Window + CL-Dual) & Text + Visual & \hspace{0.8ex}\underline{\textbf{36.01}}\textsuperscript{\textdagger} & \underline{\textbf{55.75}} & \hspace{0.8ex}\underline{\textbf{51.56}}\textsuperscript{\textdagger} & \hspace{0.8ex}\underline{\textbf{42.25}}\textsuperscript{\textdagger} & \underline{\textbf{51.91}} & \hspace{0.8ex}\underline{\textbf{49.35}}\textsuperscript{\textdagger} \\
\specialrule{.1em}{.05em}{.05em}
\end{tabular}
}
\caption{\label{tab:res_cross} Results on BBC and Behance corpora for cross-domain evaluation. \textbf{Bolded} and \underline{underlined} results indicate the best
performance across all comparisons and within their own section. $\dagger$ indicates results applied contrastive domain adaptation which are significantly different ($p<0.05$) from \textit{Ours (Window)}.}
\end{table*}

\vspace{1ex}
\noindent
\textbf{Cross-Domain Segment Inference: }Table~\ref{tab:res_cross} compares the performance of the baselines and our proposed segmenter pre-trained on YouTube w/ or w/o applying our long video adaptation strategy to two challenging long video corpora.
To better investigate the effectiveness of the two contrastive objectives associated with long video adaptation, Table~\ref{tab:res_cross} also shows the results by adding each objective (CL-Cross/CL-Intra) individually. 
The primary takeaway is that our long video adaptation strategy (Window + CL-Dual) significantly and consistently improves the performance achieved by the segmenter initially trained on short YouTube videos (Plain). 
In more detail, we can observe that utilizing the sliding window inference strategy (Plain $\rightarrow$ Window) already yields noticeable performance gains on both corpora.
Furthermore, adding contrastive learning objectives can further boost accuracy on two long video corpora. The greatest enhancement is seen when the model is tuned with dual-contrastive losses.

\vspace{-1ex}
\section{Conclusion and Future Work}
We present a multi-modal video topic segmentation model accepting both video transcripts and frames as input. 
Further, we propose a novel unsupervised domain adaptation strategy 
empowered by a dual-contrastive learning framework to generalize our model pre-trained on short videos to longer videos with more complex content 
and subtler visual changes. 
Experiments on two settings (intra-domain and cross-domain segment inference) show that (1) our system achieves the 
SOTA performance on a 
newly collected YouTube corpus consisting of large-scale but short videos; (2) When we apply our 
long video adaption strategy, 
the model for short videos 
achieves better performance when transferred to two domains comprising long (live-)stream videos. 

%
%
%
\bibliographystyle{splncs04}
\bibliography{sample_base}

\end{document}